\documentclass[conference]{IEEEtran}
\IEEEoverridecommandlockouts
\usepackage{cite}
\usepackage{amsmath,amssymb,amsfonts}
\usepackage{algorithmic}
\usepackage{graphicx}
\usepackage{textcomp}
\usepackage{xcolor}
\usepackage{comment}
\def\BibTeX{{\rm B\kern-.05em{\sc i\kern-.025em b}\kern-.08em
    T\kern-.1667em\lower.7ex\hbox{E}\kern-.125emX}}

\begin{document}

\makeatletter
\newcommand{\linebreakand}{%
  \end{@IEEEauthorhalign}
  \hfill\mbox{}\par
  \mbox{}\hfill\begin{@IEEEauthorhalign}
}
\makeatother

\title{Investigating the Impact of a Dual Musical Brain-Computer Interface on Interpersonal Synchrony: A Pilot Study}

\author{\IEEEauthorblockN{Anita Vrins}
\IEEEauthorblockA{\textit{dept. of Cognitive Science and AI} \\
\textit{Tilburg University}\\
Tilburg, Netherlands \\
}
\and
\IEEEauthorblockN{Ethel Pruss}
\IEEEauthorblockA{\textit{dept. of Cognitive Science and AI} \\
\textit{Tilburg University}\\
Tilburg, Netherlands \\
}
\and
\IEEEauthorblockN{Caterina Ceccato}
\IEEEauthorblockA{\textit{dept. of Cognitive Science and AI} \\
\textit{Tilburg University}\\
Tilburg, Netherlands \\
}

\linebreakand
\IEEEauthorblockN{Jos Prinsen}
\IEEEauthorblockA{\textit{dept. of Cognitive Science and AI} \\
\textit{Tilburg University}\\
Tilburg, Netherlands \\
}

\and
\IEEEauthorblockN{Maryam Alimardani}
\IEEEauthorblockA{\textit{dept. of Cognitive Science and AI} \\
\textit{Tilburg University}\\
Tilburg, Netherlands \\
}
}

\maketitle

\begin{abstract}

This study looked into how effective a Musical Brain-Computer Interface (MBCI) can be in providing feedback about synchrony between two people. Using a double EEG setup, we compared two types of musical feedback; one that adapted in real-time based on the inter-brain synchrony between participants (Neuroadaptive condition), and another music that was randomly generated (Random condition). We evaluated how these two conditions were perceived by 8 dyads (n = 16) and whether the generated music could influence the perceived connection and EEG synchrony between them. The findings indicated that Neuroadaptive musical feedback could potentially boost synchrony levels between people compared to Random feedback, as seen by a significant increase in EEG phase-locking values. Additionally, the real-time measurement of synchrony was successfully validated and musical neurofeedback was generally well-received by the participants. However, more research is needed for conclusive results due to the small sample size. This study is a stepping stone towards creating music that can audibly reflect the level of synchrony between individuals.
\end{abstract}


\section{Introduction}

Interpersonal synchrony, the temporal coordination of behaviors among individuals, plays a significant role in shaping social interactions, serving as a predictor for the development of affiliative perception between interacting partners \cite{chetouani2017interpersonal,Hove2009}. Further, interpersonal synchrony fosters prosocial behavior in adults, promoting cooperative and altruistic tendencies \cite{Cirelli2014}. This coordinated alignment of actions and behaviors is relevant across various social contexts, including activities such as music performances and collaborative tasks. It gives rise to a sense of shared experiences and enhanced social cohesion. Moreover, interpersonal synchrony is recognized as a critical component in early development \cite{Markova2019}. Infants and their caregivers achieve synchrony through entrainment to communicative rhythms, facilitating effective communication and nurturing social and emotional development \cite{Markova2019}.

Similarly, inter-brain synchrony, observable in both cooperative and competitive interactions, has been a focus of research. Studies underscore the role of musical rhythms in fostering interpersonal synchrony, enhancing the formation of social bonds and cooperation \cite{Hove2009,Hu2018}. Notably, akin to behavioral synchrony, neural synchrony has been found to promote prosocial behavior and strengthen affiliation between interacting individuals \cite{Cacioppo2014,Cirelli2014}. These findings suggest that neurofeedback techniques focusing on inter-brain synchrony hold potential for applications in team building and the promotion of cooperative behavior. Although there is a growing body of research exploring the potential of neurofeedback for synchrony \cite{Saul2022,chen2021hybrid,dikker2019using}, empirical studies investigating the effects of such feedback, particularly in the context of musical feedback, remain limited. Therefore, further investigations are necessary to explore and evaluate the impact of neurofeedback techniques based on inter-brain synchrony, with specific attention to the influence of musical feedback. Addressing this research gap would provide a deeper understanding of the potential benefits and practical implications of utilizing neurofeedback to promote synchrony.

To bridge this research gap and deepen our understanding of the potential benefits and practical implications of neurofeedback in promoting synchrony, this study focused on the role of interpersonal synchrony as a critical component in enhancing social bonding and cooperation, and specifically investigated the potential of musical neurofeedback to amplify this synchrony. Our experimental approach engaged unfamiliar dyads in a five-minute eye contact task, with music, generated in real-time based on the synchrony between their neural signals. The effectiveness of this approach in fostering interpersonal synchrony is validated through comparison with a control group who listened to randomly generated music. To compare the two groups, we employed both objective measures, derived from electroencephalography (EEG) data, and subjective measures, incorporating self-reported assessments and semi-structured interviews from the participants. With this comprehensive approach, our study endeavors to illuminate the potential impact and wider applicability of musical neurofeedback in enriching interpersonal synchrony, thereby influencing social dynamics and interactions.

\section{Literature Review}
Inter-Subject Synchronization (ISS) has been widely explored within the realm of shared musical experiences, with functional magnetic resonance imaging (fMRI) studies evidencing ISS during simultaneous music listening \cite{abrams2013inter}. For instance, in a study by Abrams et al. \cite{abrams2013inter}, participants listened to classical symphonies under three conditions -Natural Music (unaltered original piece), Phase-Scrambled (random phase shifts added to each frequency component), and Spectrally Rotated (modified pitch)— while their brain activity was recorded. Findings indicated significant inter-brain synchrony in the Natural Music condition, particularly within the right parietal lobe, suggesting that collective music listening can induce inter-brain synchrony.

Inter-brain synchrony has also been observed in the context of eye contact and musical engagement \cite{abrams2013inter,Zamm2021,luft2021social}. Hyperscanning studies have demonstrated this synchronization during eye contact \cite{hirsch2017frontal,koike2015hyperscanning,saito2010stay}. Notably, the Hybrid Harmony system by Chen et al. \cite{chen2021hybrid} used EEG-derived Phase Locking Values (PLV) to visualize and sonify real-time synchrony between dyads in an immersive environment, finding correlations between personal distress, social closeness, and inter-brain synchrony. However, the potential for motor artifacts during the experiments raised concerns regarding data fidelity and natural interactions.

Similar to Hybrid Harmony, we previously introduced a Musical Brain-Computer Interface (MBCI), named BrainiBeats  \cite{ceccato}, which connected two users to a live music generator through their EEG signals. However, unlike Hybrid Harmony, the BrainiBeats system extracted several features from the dual EEG signals to generate music. Specifically, three primary EEG features —signal amplitude, Frontal Alpha Asymmetry (FAA), and PLV— were extracted and mapped to musical elements such as pitch, mode, and synchrony to reflect dyads' emotions and synchrony. It is worth noting that the BrainiBeats system was only validated with one dyad, i.e., the experimenters, leaving questions about the generalizability of the system to untrained listeners \cite{ceccato}.

Despite the increasing attention on interpersonal and inter-brain synchrony and its influence on social affiliation, current research falls short in investigating the impact of interactive musical neurofeedback systems on perceived interpersonal synchrony. Moreover, there is a need to validate whether the synchrony features manifested in the generated music are audibly distinguishable and accurately represent different synchrony states.

In light of these gaps, this study aimed to expand previous work by Ceccato et al. \cite{ceccato} and posed the following research questions:

RQ1: Can the musical feedback generated by the BrainiBeats dual BCI system enhance synchrony between dyad members?

RQ2: Is there a correlation between a dyad's EEG synchrony (as measured by PLV) and their subjectively perceived interpersonal synchrony?

RQ3: Can dyad members accurately identify the levels of synchrony reflected in the musical feedback?

\section{Methods}

\begin{figure*}[htbp]
\centerline{\includegraphics[width=0.8\textwidth]{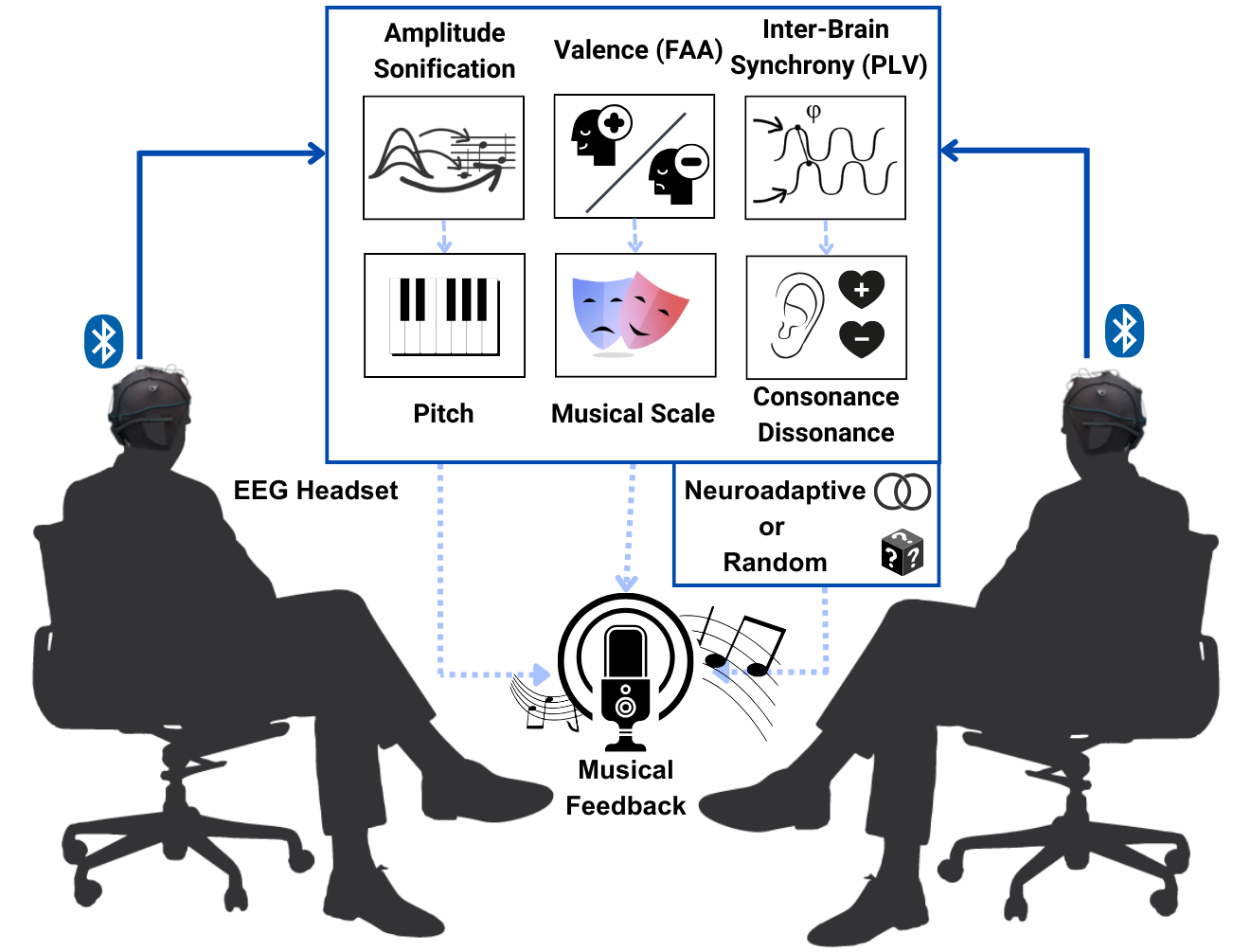}}
\caption{The dual BCI System used by two users to generate neuroadaptive musical feedback based on their inter-brain synchrony. A higher level of synchrony between dyad members was translated into more consonant (i.e., more pleasant-sounding) music and a low synchrony was translated into dissonant (i.e., more jarring) music. The system is an updated version of the BrainiBeats system from
\cite{ceccato}.}
\label{fig:bciloop}
\end{figure*}

\subsection{Study design}
This study utilized a between-group design with a total of 8 dyads (n = 16), all of whom were university students (6 females  10 males, age range of 18-24). To mitigate the influence of emotional closeness and PLV values, as indicated by Chen et al. (2021), dyad partners who were unfamiliar with each other were paired. Half of the dyads were assigned to the \emph{Neuroadaptive music} condition, where inter-brain synchrony features were measured and reflected in the generated music. The remaining dyads were assigned to the \emph{Random music} condition, where musical features associated with synchrony were randomized.
Two questionnaires were administered: one before the experimental interaction and one afterwards. All participants were provided with an information letter and consent form prior to the experiment.

\subsection{Musical BCI System}
The experimental system used in this study was an updated version of the BrainiBeats system developed by Ceccato et al. \cite{ceccato}. Figure \ref{fig:bciloop} illustrates the updated mapping scheme for converting EEG signals from two people to music. For data collection, two modified 8-channel Unicorn Hybrid Black EEG headsets with the universal 10/20 electrode placement (Fz, F3, F4, C3, Cz, C4, Pz, and Oz) were utilized to capture neurophysiological features for sonification (PLV, FAA and EEG amplitudes; see \cite{ceccato} for a more detailed description of signal processing, feature extraction and sonification).

Several modifications were implemented to the system reported in \cite{ceccato}, including updates to the amplitude-to-pitch mapping. Previously, the mapping involved a scale-free conversion of amplitude to MIDI notes. However, to improve the quality and pleasantness of the generated music, the pitch values were adjusted to the closest stable-sounding notes on the chosen scale. The scale was determined by valence (FAA) where negative valence was represented by the minor scale and positive valence by the major scale. As valence was measured for both participants, the scale determination alternated between participants for each note. While the mapping of consonance (high synchrony) and dissonance (low synchrony) stayed the same, in this version, a drone note was used to achieve the effect: a sustained note was played in the background, which either harmonized with the given scale or did not, depending on the level of inter-brain synchrony (PLV). Furthermore, adjustments were made to the instrumental loop in the Sonic Pi music generation software to enhance overall music quality and make the musical features more distinguishable. Finally, the live audio stream \cite{brainihacks-2023} was played from the main processing computer.

\subsection{Experimental protocol}
The procedure started with participants providing consent. Following that, participants completed an initial questionnaire asking about their demographic data, their initial impressions of their partner and their level of synchrony before starting the experiment. To minimize proximity effects in this stage, we ensured limited visibility between the dyads by seating them in separate cubicles or positioning them out of each other's view. After placing the EEG headsets and obtaining a baseline recording (1 minute of looking at a neutral object) from both participants, they were asked to sit facing each other (approximately 1.5 meters apart) and maintain eye contact. This phase lasted 5 minutes during which participants listened to the generated music that was either based on the dyad's EEG activity (Neuroadaptive condition) or was produced randomly (Random condition). Post-experiment questionnaires were then filled out by each participant, gathering feedback on the music and their perceived synchrony with each other, finally, a shared interview was conducted with the dyad about their experience.

\subsection{Evaluation measures}
\subsubsection{Phase Locking Value (PLV)}
As a measure of EEG-based interpersonal synchrony, the Phase Locking Value (PLV) between the two participants (averaged over all electrode pairs) was calculated live by the BCI system and stored as an objective measure of synchrony during the baseline phase (no eye contact) and during the eye-contact phase. For details of the PLV calculation, see Ceccato et al. \cite{ceccato}. To compare the effect of experimental conditions, the change in PLV from the baseline to the eye-contact phase ($PLV_{eye contact}$ -- $PLV_{baseline}$) was computed per dyad and the outcome was compared between the two conditions using statistical tests.

\subsubsection{Questionnaires}
Questionnaires filled before and after the experiment assessed participants' perceived interpersonal synchrony \cite{popovic-2003}. In the post-questionnaire, additional questions were included about participant's perception of the music in terms of enjoyment as well as the representation of their mood, mental state and synchrony with the other participant. All measures used a 5 or 7-point Likert scale.

\subsubsection{Interviews}
The post-experiment interviews involved both dyad members to gather their shared insights. The interviews asked whether the dyad enjoyed the music and if they felt that the music represented their mental states. At the end of the interview, participants were informed about the two conditions one involving interpersonal synchrony as a music element, and the other being randomized in that aspect. Participants were then asked to guess which condition they were assigned to based on this information. Finally, we inquired about potential improvements to the system, encompassing both the experimental setup and music generation. 

\subsection{Statistical analysis}

Given the small sample size and non-normal data distribution, non-parametric statistical methods (Wilcoxon Signed Rank tests and Spearman's Rank-Order Correlation) were used for data analysis. For the post-experiment interview, audio transcripts were analyzed using Nvivo 1.7 \cite{lumivero-2023}.

\section{Results}

\subsection{The impact of the BCI system on interpersonal synchrony}

The system's ability to measure and reflect interpersonal synchrony (inter-brain PLV) on musical feedback was validated using two methods: 1) comparing the change in subjective and objective synchrony indicators after the intervention (5 minutes of eye contact) that was expected to enhance synchrony  \cite{hirsch2017frontal,koike2015hyperscanning,saito2010stay}, and 2) analyzing the correlation between subjective synchrony scores and system-measured PLV during the eye-contact (i.e., musical feedback) phase.

\subsubsection{Change in subjective and objective synchrony}

Baseline-corrected Phase-Lock Value (PLV) and change in subjective synchrony were compared between Neuroadaptive and Random Feedback conditions using Wilcoxon signed-rank tests (Figure \ref{fig:delta_synchrony}). Notably, the Neuroadaptive feedback condition showed a significantly higher increase in baseline-corrected PLV ($Z$ = 1.73, $p$ = .042; $Md$ = .46) than Random Feedback ($Md$ = -.02). The subjective synchrony changes were not significantly different between the two conditions although a trend was observed ($Z$ = -1.48, $p$ = .070; $Md$ = 2.0 in both conditions).

\begin{figure}[htbp]
\centerline{\includegraphics[width=0.5\textwidth]{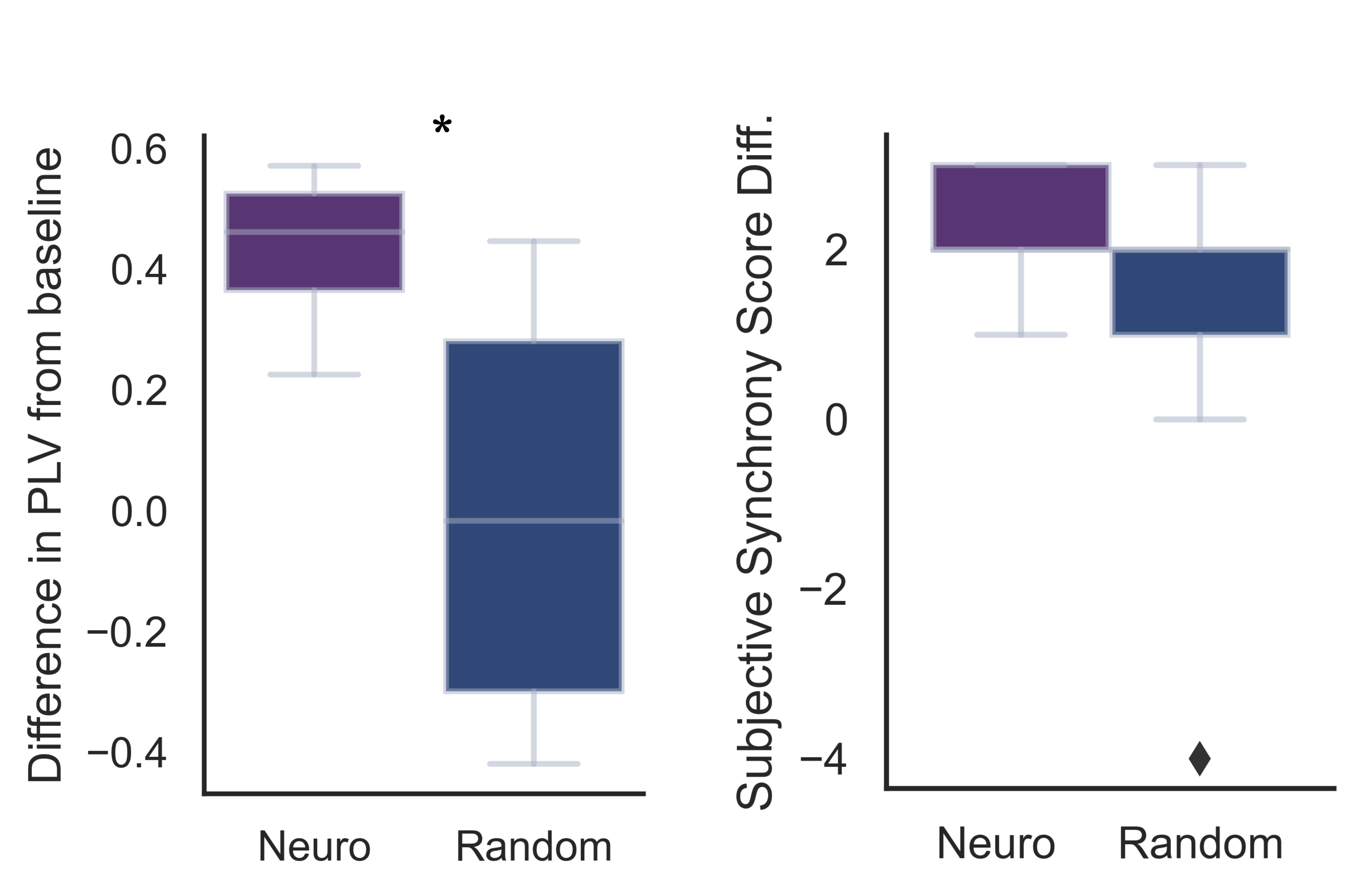}}
\caption{The change in inter-brain PLV and subjective synchrony from the baseline to the eye-contact phase compared between Neuroadaptive and Random feedback conditions. (* $p \leq 0.05$).}
\label{fig:delta_synchrony}
\end{figure}

\subsubsection{Correlation between PLV and subjective synchrony in each condition}

A significant positive correlation ($r_s$(14) = .73, $p$ = .001) was observed between subjective synchrony and PLV during the eye-contact phase only in the Neuroadaptive feedback condition (Figure \ref{fig:correlation_separate}), but this was not the case in the Random condition ($r_s$(14) = .03, $p$ = .92). Please note that the same PLV value obtained during the eye-contact phase has been assigned to both dyad members in this analysis. 

\begin{figure}[htbp]
\centerline{\includegraphics[width=0.5\textwidth]{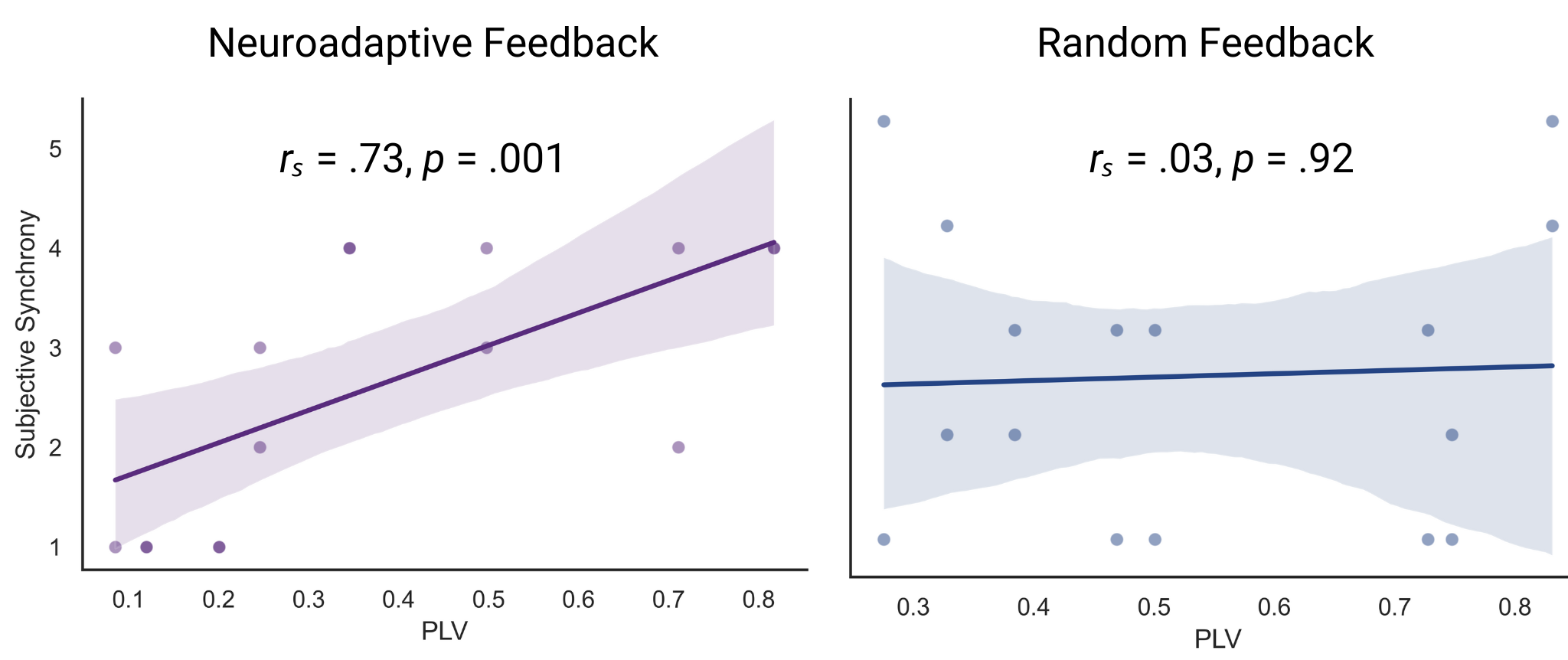}}
\caption{Relationship between subjective synchrony and PLV changes per feedback condition.}
\label{fig:correlation_separate}
\end{figure}

\subsection{Participants' perception of the musical feedback}

\subsubsection{Questionnaire Insights}

Participants' perceptions of musical feedback in Neuroadaptive and Random Feedback conditions were compared across four dimensions: enjoyment, representation of mental state, mood, and synchrony using Wilcoxon signed-rank tests (Figure \ref{fig:music}). Significant differences between the conditions were found in enjoyment ($Z$ = 1.91, $p$ = .028; $Md_{Neuro}$ = 3.5, $Md_{Random}$ = 2.0) and perceived mental state representation ($Z$ = 1.69, $p$ = .046; $Md_{Neuro}$ = 3.0, $Md_{Random}$ = 2.5). No significant difference was noted in music representation of mood ($Z$ = 1.42, $p$ = .078; $Md_{Neuro}$ = 3.0, $Md_{Random}$ = 2.0) and synchrony ($Z$ = .49, $p$ = .313; $Md_{Neuro}$ = 4.0, $Md_{Random}$ = 3.5).

\begin{figure}[htbp]
\centerline{\includegraphics[width=0.5\textwidth]{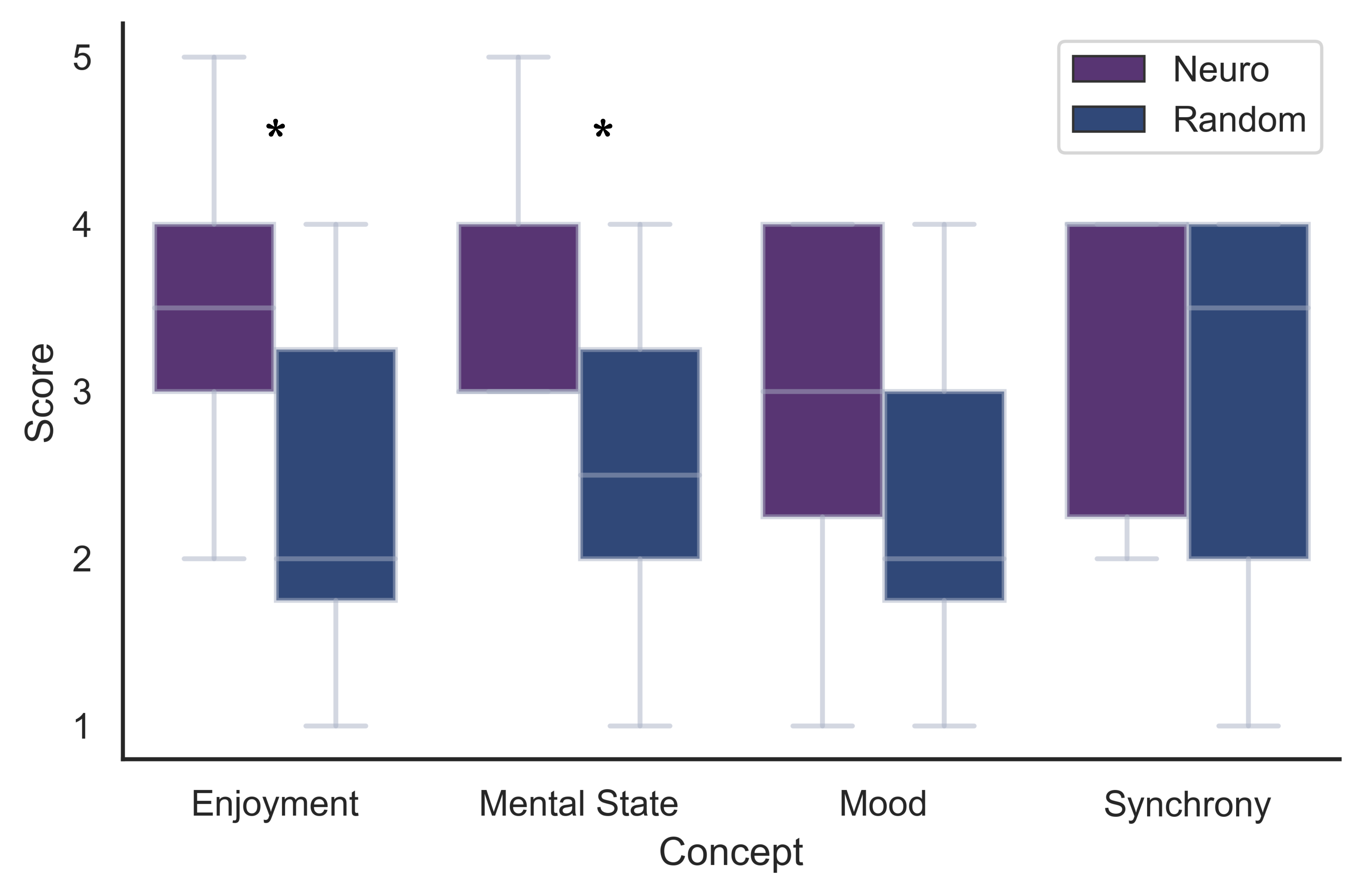}}
\caption{Participants' self-reported perception of the musical feedback generated by the BCI system in the Neuroadaptive and Random feedback conditions. (* $p \leq 0.05$).}
\label{fig:music}
\end{figure}

\subsubsection{Semi-structured interviews}

The interview was conducted only with participants who were blind to the design of the experiment (n = 12). Most were able to correctly identify the feedback condition they were assigned to (9/12). The majority found the music calming (10/12), but those in the Random Feedback condition also likened it to white noise or rhythmic noise (6/6 in the Random condition). An increase in variety of the music was suggested as the main point of improvement (10/12), as the presented version was perceived as quite monotonous. According to the users, any changes to the music were very subtle and barely noticeable.

\section{Discussion}

This study aimed to evaluate if a dual Musical Brain-Computer Interface (MBCI) that generated music based on inter-brain synchrony between two users \cite{ceccato}, could enhance the dyad's neural and subjective synchrony compared to when a dyad listened to a randomly generated musical feedback. Despite the study's small sample size, initial insights were gained.

Our findings suggest that the proposed MBCI system can reliably measure and influence interpersonal synchrony with the neuroadaptive music inducing a significantly higher increase in inter-brain Phase-Locking Values (PLV) compared to the randomly generated music. Additionally, a positive correlation between subjective synchrony and PLV was found when the participants listened to the neuroadaptive musical feedback during the eye-contact phase, but not in the case of randomly generated music. These results replicate findings from past research that suggested increased synchrony between two people during eye contact \cite{hirsch2017frontal,koike2015hyperscanning,saito2010stay} and further validate the role of neurofeedback (here in form of music) during such tasks in deepening people's understanding of neural synchrony. Yet, robust conclusions necessitate larger samples and stringent control for confounding variables.



In terms of perception of music, participants reported enjoying music in the Neuroadaptive feedback condition more and believed it better represented their mental state; semi-structured interviews corroborated the ability to distinguish between Neuroadaptive and Random feedback. However, no significant differences were noted in subjectively perceived synchrony, indicating potential limitations in interpreting what musical feedback represents. Semi-structured interviews revealed similar themes: while participants could recognize which condition they were in, they were confused by what the music represented and suggested more pronounced musical features to identify synchrony, as well as more variation in the music in order to make it more enjoyable and engaging.

These initial findings illuminate the potential of a Musical BCI system for enhancing and providing feedback on interpersonal synchrony. Based on results from Hybrid Harmony by \cite{chen2021hybrid}, we controlled this pilot experiment for effects of social familiarity between dyads by pairing together strangers only. The novelty of our experiment is that we measured a baseline of both the subjective first impressions of the dyad partners as well as a baseline for PLV when the dyads were outside of each others line of sight. Therefore we show the effect our Musical BCI had specifically on pairs of strangers. Possible use-cases of our musical BCI system could be in promoting social relationships between strangers for example in team building activities in professional settings or to promote social bonding among peers for example in classroom settings. 


\section{Conclusion}

This preliminary investigation provides initial support for the potential of a dual Musical Brain-Computer Interface (MBCI) in enhancing and providing feedback on interpersonal synchrony. In addition to validating the system’s measurement of synchrony, the research found that neurofeedback through adaptive music could potentially elevate synchrony levels between strangers, and participants were generally able to discern between neuroadaptive and randomly generated music, indicating an increased sense of enjoyment and representation of their mental state in the neuroadaptive condition.

Despite these promising insights, caution should be exercised in interpreting the results due to the small sample size and potential confounding effects. Notably, improving participants' understanding of the musical feedback could be beneficial, with suggestions including clearer musical synchrony cues and greater feedback variation. Future research should build on these findings, refine the musical feedback system, and incorporate larger sample sizes for more definitive results. 

\bibliographystyle{IEEEtran}
\bibliography{IEEEabrv,main}

\end{document}